\documentstyle[preprint,aps]{revtex}
\tighten

\begin{document}
\draft
\preprint{Preprint Numbers:
          \parbox[t]{50mm}{FSU-SCRI-93-132\\ ADP-93-223/T140}}
\title{Medium Modifications to the $\omega$-Meson Mass\\
       in the Walecka Model}
\author{H.-C. Jean\footnotemark[1]\footnotemark[2],
        J. Piekarewicz\footnotemark[2], and
        A. G. Williams\footnotemark[1]\footnotemark[2]\footnotemark[3]
        \vspace*{2mm}}
\address{\footnotemark[1]Department of Physics,
         Florida State University, Tallahassee, FL 32306\\*[2mm]
       \footnotemark[2]Supercomputer Computations Research Institute,
         Florida State University,\\
         Tallahassee, FL 32306\\*[2mm]
       \footnotemark[3]Department of Physics and Mathematical Physics,
         University of Adelaide,\\
         S. Aust. 5005, Australia}
\date{\today}
\maketitle

\begin{abstract}
We calculate the effective mass of the $\omega$ meson in nuclear
matter in a relativistic random-phase approximation to the Walecka
model. The dressing of the meson propagator is driven by its coupling
to particle-hole pairs and nucleon-antinucleon ($N\bar{N}$)
excitations.  We report a reduction in the $\omega$-meson mass of
about 170~MeV at nuclear-matter saturation density. This reduction
arises from a competition between the density-dependent
(particle-hole) dressing of the propagator and vacuum polarization
($N\bar{N}$ pairs). While density-dependent effects lead to an
increase in the mass proportional to the classical plasma frequency,
vacuum polarization leads to an even larger reduction caused by the
reduced effective nucleon mass in the medium.
\end{abstract}

\narrowtext

\section{Introduction}
\label{secintro}

Understanding the role played by the nuclear medium in modifying
hadronic properties is one of the most interesting and challenging
problems facing nuclear physics today. For example, the
spin-independent, (or old) European Muon Collaboration (EMC)
experiment revealed a medium-modified electromagnetic coupling of the
nucleon relative to its free-space value. Trying to explain the origin
of this modification, in terms of, either, conventional nuclear
physics effects (e.g., binding energy, Fermi motion, correlations) or
more exotic mechanisms (e.g., nucleon swelling), is still the source
of considerable debate.

Also very interesting is the study of the modification of meson
properties in the medium. Indeed, many interesting phenomena in finite
nuclei have been attributed to an in-medium reduction of the mass of
the rho meson\cite{brown90}. These phenomena include the lack of an
enhancement in the ratio of the spin-longitudinal to spin-transverse
responses measured in quasielastic $(\vec{p},\vec{n})$
scattering\cite{brown93}, the enhancement of the $K^{+}$-nucleon
interaction in the medium\cite{brown88}, and the behavior of certain
spin observables measured in inelastic $(\vec{p},\vec{p}\:^{'})$
transitions\cite{steph88}.

In the medium, a meson gets modified due to its coupling to nuclear
excitations. This modification is contained in the meson self-energy
whose imaginary part is a physical observable characterizing the
linear response of the nuclear system to an external probe. To date,
much work has been done (experimentally and theoretically) in
understanding the response of the nuclear system in the spacelike
region (i.e., $q_{\mu}^{2} <0$). All of the information gathered so
far about the response of the nuclear system to a variety of probes
(e.g., $e^{-}$, $\pi$, $K^{+}$, $N$), and for a variety of kinematical
conditions (covering the inelastic, giant-resonance, quasielastic
regions) can only reveal the nature of the nuclear response in the
spacelike region. In these experiments the coupling of the probe to
timelike excitations can only occur virtually. It therefore becomes
very interesting to study the behavior of the meson self energy in the
timelike region. This could be done, for example, in colliding
$(e^{+}e^{-})$ experiments and relativistic heavy-ion collisions.
Alternatively, it can be studied by directly measuring the
medium-dependence of the decay of the meson into lepton pairs. Indeed,
a proposal has been put forward to measure (at CEBAF) the nuclear-mass
dependence of vector mesons by detecting lepton pairs\cite{cebaf89}.

The dependence of meson properties on the density of
the nuclear medium is far from understood. In particular, predictions
for the shift in the value of the $\omega$ meson at normal
nuclear-matter density range anywhere from $-100$ to +100~MeV. These
predictions are based on a variety of models that include quantum
hadrodynamics (QHD)\cite{chin77}, Nambu-Jona-Lasinio\cite{hosaka90},
and QCD sum rules (QSR)\cite{hatlee92,koike93}.

In this paper we attempt a more detailed analysis and so elect to use
the simplest version of QHD, namely, the Walecka model, to study
medium modifications to the $\omega$-meson propagator in the timelike
region. The Walecka model is a strong-coupling renormalizable field
theory of nucleons interacting via the exchange of (isoscalar) scalar
($\sigma$) and vector ($\omega$) mesons\cite{wal74,serwal86}. The
model has already been used extensively in calculations of nuclear
matter and finite nuclei. The saturation of nuclear matter and the
strong spin-orbit splitting observed in finite nuclei were among the
first successes of the model\cite{wal74,serwal86,horser81,horser84}.
More recently, the model has been used to calculate such diverse
topics as collective modes in nuclear matter\cite{limhor88},
isoscalar magnetic moments\cite{mcneil86,furser87}, and
electroweak\cite{wehbec88,horpie89,horpie93} and hadronic responses
from finite nuclei\cite{korma93} with considerable success.

In spite of these successes, the theory remains (practically) untested
in the timelike region, which holds special relevance for a model that
incorporates negative-energy degrees of freedom from the outset. In
this calculation we will show that the in-medium shift in the value of
the $\omega$-meson mass arises from a sensitive cancellation between
two competing effects. On the one hand, the (virtual) coupling of the
meson to particle-hole excitations leads to an increase in the value
of the mass. This is consistent with the result reported by Chin using
an approximation to QHD that ignored the coupling of the $\omega$
meson to $N\bar{N}$ excitations (vacuum polarization)\cite{chin77}.
By also including vacuum polarization in the present calculation, we
believe that we have performed a more consistent field-theoretical
calculation. We have found that $N\bar{N}$ excitations generate a
shift in the value of the $\omega$-meson mass proportional to the
shift of the nucleon mass in the medium. Since in QHD the scalar field
is responsible for a reduction of the nucleon mass, $N\bar{N}$
excitations (by themselves) give rise to a reduction of the
$\omega$-meson mass. For values of the model parameters consistent
with the description of nuclear matter at saturation, we have found
that vacuum polarization overwhelms the corresponding
density-dependent contribution, and, ultimately, leads to a reduction
of the $\omega$-meson mass in the medium.

We have organized the paper as follows. In Sec.~\ref{secformal} we
present the formalism needed to calculate the self-energy corrections
to the propagation of the $\omega$ meson through the nuclear medium.
In Sec.~\ref{secres} we present results for the effective mass of the
$\omega$ meson with special emphasis on the competition between the
density-dependent dressing of the meson and vacuum polarization.
Finally, in Sec.~\ref{secconcl} we offer our conclusions and outlook
for future work.

\section{Formalism}
\label{secformal}

In this section we calculate the self-energy corrections to the
$\omega$-meson propagator in nuclear matter. The dressed meson
propagator will be calculated to one-loop order by solving Dyson's
equation in a nuclear-matter ground state obtained from using a
relativistic mean-field approximation to the Walecka model.

The Walecka model (or QHD-I) is a renormalizable, relativistic
quantum field theory of the nuclear system which involves an explicit
description of the nucleon ($N$) and meson ($\sigma,~\omega$) degrees
of freedoms\cite{wal74,serwal86}. The Lagrangian density for the
Walecka model is
\begin{eqnarray}
  {\cal L} & = & \bar{\psi} [ \gamma_\mu (i\partial^\mu-g_{v} V^\mu)
    - (M-g_s\phi)]\psi\nonumber\\
  && + {1\over2} (\partial_\mu\phi\partial^\mu\phi-m_s^2\phi^2)
    \nonumber\\
  && - {1\over4}F_{\mu\nu}F^{\mu\nu} + {1\over2}m_{v}^{2}V_\mu V^\mu
    + \delta{\cal L} \;,
\end{eqnarray}
where $\psi$ is the baryon field with mass $M$, $\phi$ is the neutral
scalar-meson ($\sigma$) field with mass $m_s$, $V^\mu$ is the neutral
vector-meson ($\omega$) field with mass $m_{v}$, and
$F^{\mu\nu} \equiv \partial^\mu V^\nu - \partial^\nu V^\mu$.
The term $\delta{\cal L}$ contains renormalization counterterms. In
this model, nucleons interact via the exchange of isoscalar mesons
with the coupling of the scalar field $\phi$ to the baryon scalar
density $\bar\psi\psi$, and the vector field $V^\mu$ to the conserved
baryon current $\bar\psi\gamma_\mu\psi$ introduced through minimal
substitution.

Since the exact solutions to the field equations are unknown (and
perhaps unattainable) we resort to a mean-field approximation in the
usual way. In a mean-field approximation, one replaces the meson field
operators by their (classical) ground-state expectation values:
\begin{mathletters}
\begin{eqnarray}
  \phi  & \rightarrow & \langle\phi\rangle \equiv \phi_0 \;,\\
  V^\mu & \rightarrow & \langle V^\mu \rangle \equiv g^{\mu 0}V^0 \;.
\end{eqnarray}
\end{mathletters}
The mean-field equations can now be solved exactly with the solution
becoming increasingly valid with increasing baryon density.
Traditionally, the mean-field equations have been solved in two
approximations. In the mean-field theory (MFT) one calculates a baryon
self-energy which is generated by the presence of all the nucleons in
the occupied Fermi sea. The effect of the (infinite) Dirac sea is,
however, neglected. In contrast, in the relativistic Hartree
approximation (RHA) one includes the contribution to the baryon
self-energy arising from the occupied Fermi sea as well as from the
full Dirac sea. As a consequence, the baryon self-energy diverges in
the RHA and must be renormalized.

In both approximations, the mean scalar-meson field $\phi_0$ is
responsible for a (downward) shift of the nucleon mass $M^{*}$ in the
nuclear medium relative to its free-space value $M$. In contrast to
the ground-state expectation value of the vector field, which is fully
determined by the conserved baryon density,
\begin{equation}
  g_{v} V^{0}  = {g_{v}^{2} \over m_{v}^{2}}
    \langle \bar\psi\gamma^{0}\psi \rangle
    = {g_{v}^{2} \over m_{v}^{2}} \rho_{B}  \;,
\end{equation}
the expectation value of the scalar field (and consequently the
effective mass) is a dynamical quantity that must be determined
self-consistently from the equations of motion
\begin{equation}
  M-M^{*} = g_{s}\phi_{0} = {g_{s}^{2} \over m_{s}^{2}}
    \langle \bar\psi\psi \rangle(M^{*})
    = {g_{s}^{2} \over m_{s}^{2}} \rho_{S}(M^{*}) \;.
\end{equation}
There are five parameters to be determined in the model. The nucleon
mass and the $\omega$-meson mass are fixed at their physical values
($M=939$~MeV and $m_{v}=783$~MeV, respectively). The other three
parameters must be fixed from physical observables. For example, the
ratios of the coupling constant to the meson mass
[$C_{s}^{2}=g_{s}^{2}M^{2}/m_{s}^{2})$ and
 $C_{v}^{2}=g_{v}^{2}M^{2}/m_{v}^{2})$]
can be chosen to reproduce the bulk binding energy (15.75~MeV) and
density (corresponding to a Fermi momentum of
$k_F=1.3\mbox{ fm}^{-1}$) of nuclear matter at saturation. Finally,
the mass of the scalar meson is adjusted to reproduce the
root-mean-square radius of $^{40}$Ca. These parameters along with the
effective nucleon mass $M^{*}$ at saturation density are listed in
Table~\ref{tbl.para}.

In order to compute the meson propagators in the nuclear medium we
solve Dyson's equation in the ring, or random-phase (RPA),
approximation\cite{fetwal71}. This approximation is characterized by
an infinite summation of the lowest-order proper polarization. In a
relativistic theory of nuclear structure the polarization insertion,
or meson self-energy, describes the coupling of the meson to two kinds
of excitations; the traditional particle-hole pairs and
nucleon-antinucleon ($N\bar{N}$) excitations. In the nuclear medium
real particle-hole excitations can be produced only if the
four-momentum carried by the meson is spacelike ($q_{\mu}^{2}<0$). In
contrast, real $N\bar{N}$ pairs can be excited only in the timelike
region ($q_{\mu}^{2}>0$).

To date, most of the relativistic RPA studies of the nuclear system
have been carried out in a  kinematical domain considerably different
from the one relevant to the present analysis. These studies have
investigated the response of the nuclear system to a variety of
probes. In all of these cases the four-momentum transferred to the
nucleus (and hence carried by the mesons) was constrained to the
spacelike region. Hence, $N\bar{N}$ pairs could only be virtually
excited.In the present work we need to study the $\omega$-meson
self-energy in a (timelike) region around the position of the
$\omega$-meson pole $(q_{\mu}^{2}\simeq m_{v}^{2})$. This is a
kinematical region where both particle-hole as well as $N\bar{N}$
pairs can only be virtually created (at least at low density) and
where the Walecka model has been largely untested.

In the Walecka model one can not decouple the $\sigma$ meson from the
analysis of the $\omega$-meson propagator. In the nuclear medium the
$\omega$ and $\sigma$ propagators are inextricably linked because of
scalar-vector mixing. Scalar-vector mixing occurs, for example, when
a particle-hole pair becomes excited in the medium by means of a
(longitudinal) vector meson which subsequently decays into a scalar
meson. Scalar-vector mixing is a purely density-dependent effect that
generates a coupling between the Dyson's equation for the scalar- and
vector-meson propagators. The full scalar-vector meson propagator and
the lowest-order (one-loop) proper polarization can be represented by
a set of $5\times5$ matrices $D^{ab}$ and $\Pi_{ab}$, with the indices
$a$ and $b$ in the range $-1, 0, 1, 2, 3$\cite{limhor88}. In this
representation Dyson's equation becomes a (coupled) matrix equation
given by
\begin{equation}
  D = D_0 + D_0\Pi D \;,
\end{equation}
where $D_0$ represents the lowest-order meson propagator
\begin{equation}
  D_0 = \left( \begin{array}{cc}
    \Delta_0 & 0            \\
    0        & D_0^{\mu\nu} \end{array} \right) \;,
\end{equation}
written in terms of the non-interacting scalar- and vector-meson
propagators
\begin{eqnarray}
  \Delta_0(q)  &=& {1\over q_\mu^2 - m_{s}^{2} + i\eta}      \;, \\
  D_0(q)       &=& {1\over q_\mu^2 - m_{v}^{2} + i\eta}      \;, \\
  D_0^{\mu\nu} &=&
    \left( -g^{\mu\nu} + q^{\mu}q^{\nu}/m_{v}^{2} \right) D_{0}(q) \;,
\end{eqnarray}
and where $q_\mu^2\equiv q_0^2-{\bf q}^2$. Notice that since the
$\omega$ meson always couples to a conserved baryon current the
$q^{\mu}q^{\nu}$ term in $D_0^{\mu\nu}$ will not contribute to
physical quantities.

The lowest-order polarization insertions will be expressed in terms of
the self-consistent nucleon propagator. The nucleon propagator is
written as a sum of Feynman [$G_F(k)$] and density-dependent
[$G_D(k)$] contributions, i.e.,
\begin{eqnarray}
  G(k)   & \equiv & G_F(k)+G_D(k)                \;, \\
  G_F(k) & =      & (\gamma^\mu k_\mu^{*}+M^{*})
    {1\over k_\mu^{*2}-M^{*2}+i\eta}             \;, \\
  G_D(k) & =      & (\gamma^\mu k_\mu^{*}+M^{*}) \nonumber\\
  && \times {i\pi\over E^{*}({\bf k})}
    \delta \Big(k_0^{*}-E^{*}({\bf k})\Big) \theta(k_F-|{\bf k}|) \;,
\end{eqnarray}
where the momentum $k^{*\mu}$ and energy $E^{*}({\bf k})$ are defined,
respectively, by $ k^{*\mu}\equiv(k^0\!-\!g_{v}V^{0},{\bf k})$ and
$E^{*}({\bf k})\equiv\sqrt{{\bf k}^2+M^{*2}}$. The Feynman part of the
propagator has the same analytic structure as the free propagator. The
density-dependent part, on the other hand, corrects $G_{F}$ for the
presence of occupied states below the Fermi surface and vanishes at
zero baryon density. In terms of the nucleon propagator the
lowest-order scalar-scalar, vector-vector, and scalar-vector (mixed)
polarizations are given, respectively, by
\begin{mathletters}
\begin{eqnarray}
  \Pi^{s}(q)          &=& -ig^2_s \int {d^4 k\over (2\pi)^4}
    \mbox{Tr} [G(k)G(k+q)]                       \;, \\
  \Pi^{v}_{\mu\nu}(q) &=& -ig^{2}_{v} \int {d^4 k\over (2\pi)^4}
    \mbox{Tr} [\gamma_\mu G(k)\gamma_\nu G(k+q)] \;, \\
  \Pi^{m}_{\mu}(q)    &=& ig_s g_{v} \int {d^4 k\over (2\pi)^4}
    \mbox{Tr} [\gamma_\mu G(k)G(k+q)]            \;.
\end{eqnarray}
\end{mathletters}
In the above expressions the traces include isospin and we have
adopted the conventions of Ref.~\cite{bjodre65}. As in the case of the
nucleon propagator we can write the above polarization insertions as a
sum of two contributions
\begin{equation}
  \Pi^{ab}(q) \equiv \Pi^{ab}_{F}(q) + \Pi^{ab}_{D}(q) \;.
\end{equation}
The Feynman contribution to the polarization, or vacuum polarization
($\Pi^{ab}_{F}$), is a bilinear function of $G_{F}$ and describes the
self-energy corrections to the meson propagators due to their coupling
to $N\bar{N}$ excitations. The Feynman contribution to the
polarization insertion diverges and must be renormalized. The
density-dependent part of the polarization ($\Pi^{ab}_{D}$), on the
other hand, is finite and contains at least one power of $G_{D}$. The
density-dependent part of the polarization insertion describes the
coupling of the meson to particle-hole excitations. In addition, it
contains a term which has no nonrelativistic counterpart. This term
arises from the negative-energy components in the Feynman propagator
and describes the Pauli blocking of $N\bar{N}$ excitations. This term
enforces the Pauli principle by preventing the nucleon from the
$N\bar{N}$ pair (in $\Pi^{ab}_{F}$) to make a transition to an
occupied state below the Fermi surface\cite{horpie89,piekar90}.

For a mean-field ground state obtained in the MFT (no vacuum loops)
approximation, it has been shown that the consistent linear response
of the system, and hence the consistent meson self-energy, is obtained
by neglecting the Feynman part of the polarization insertion. This
consistency is reflected, for example, in the proper treatment of
(spurious) excitations associated with an overall translation of the
center of mass of the system. Notice, however, that in the MFT
approximation one retains the Pauli blocking of $N\bar{N}$ excitations
even though one is neglecting the Feynman contribution to the
polarization. Retaining the Pauli blocking of $N\bar{N}$ excitations
has been proven essential to satisfy current conservation in
calculations of the electromagnetic response of the nuclear system.
Nevertheless, one should question an approximation that retains the
Pauli blocking of an excitation that has not been put in from the
outset. If the effect from Pauli blocking of $N\bar{N}$ excitations
represents a small contribution to the overall size of the
observables, then one might be justified in making this approximation.
If, however, the observables are seen to depend heavily on this
assumption, then one would be forced to neglect this approximation in
favor of the RHA where vacuum loops are included in, both, the
description of the ground state as well as in
the linear response of the system. We now present results for the
effective mass of the $\omega$ meson in both (MFT and RHA)
approximations.

\section{Results}
\label{secres}

The effective mass of the $\omega$ meson ($m_{v}^{*}$) in nuclear
matter is obtained by finding the value of the four-momentum
$q_{\mu}^{2}$ for which the imaginary part of the propagator attains
its maximum. Alternatively, since the $\omega$ meson has a very narrow
width in the region of interest, we can find the effective mass by
searching for zeroes in the inverse propagator. In the particular case
of the transverse component of the polarization (unaffected by
scalar-vector mixing) we obtain
\begin{eqnarray}
  D_{T}^{-1}(q) & \equiv & D_{22}^{-1}(q) = D_{33}^{-1}(q) \nonumber\\
  & = & q_{\mu}^{2} - m_{v}^{2} - \Pi_{22}(q;k_{F}) = 0 \;,
 \label{dtrans}
\end{eqnarray}
where we have defined the x-axis along the direction of
three-momentum ${\bf q}$. The (transverse) effective $\omega$-meson
mass has been plotted in Fig.~\ref{figone} as a function of the Fermi
momentum (relative to its value at saturation
$k_{F}^{0}=1.30$~fm$^{-1}$) for two values of the three-momentum
transfer. These MFT results are in agreement with those published in
Ref.~\cite{limhor88} and have been included here for completeness.

One can gain some insight into the physics driving these modifications
to the $\omega$-meson mass by examining the low-density limit of these
results. By performing a low-density expansion of the transverse meson
propagator one can show that effective mass of the $\omega$ meson is
given by
\begin{equation}
  m_{v}^{*2} = m_{v}^{2} + \Omega^{2} + {\cal O}(m_{v}^{2}/4M^{2}) \;,
  \label{mvmft}
\end{equation}
where we have introduced the classical plasma frequency
\begin{equation}
  \Omega^{2} = {g_{v}^{2} \rho_{B} \over M} \;.
\end{equation}
The above set of equations indicate that the density-dependent
dressing of the $\omega$-meson propagator leads to an increase in the
mass of the $\omega$-meson in the medium which is proportional to the
classical plasma frequency. This is the (old) result obtained by Chin
in 1977, e.g., see Ref.~\cite{chin77}.

Two features of this plot are particularly noteworthy. First, the area
between the dotted lines represents the region where the imaginary
part of the transverse polarization is nonzero. Inside this region the
$\omega$ meson can ``decay'' into $N\bar{N}$ pairs. This region is
bounded from below by the curve
\widetext
\begin{equation}
  [q^{0}]_{\rm min} = \cases{
    \sqrt{{\bf q}^{2}+4M^{*2}} \;, & if $|{\bf q}| \le 2k_{F} \;;$ \cr
    \sqrt{k_{F}^{2}+M^{*2}} + \sqrt{(|{\bf q}|-k_{F})^{2}+M^{*2}} \;,
      & if $|{\bf q}| \ge 2k_{F} \;,$ \cr}
\end{equation}
\narrowtext
\noindent and from above by the curve
\begin{equation}
  [q^{0}]_{\rm max} = \sqrt{k_{F}^{2} + M^{*2}} +
                      \sqrt{(|{\bf q}|+k_{F})^{2} + M^{*2}} \;.
\end{equation}
This region is defined by imposing energy-momentum conservation for
the (on-shell) production of a $N\bar{N}$ pair with the nucleon
three-momentum constrained to be below the Fermi momentum. Note,
however, that by itself this nonzero imaginary part contributes to an
unphysical (i.e., negative) decay width for the $\omega$ meson. This
can be seen, for example, by considering the case of $N\bar{N}$ pair
production at ${\bf q}=0$ ($N\bar{N}$ pair produced with equal
magnitude but opposite direction of the three-momentum). In the
absence of Pauli blocking the threshold for pair production would
start at $q^{0}=2M^{*}$ (nucleon and antinucleon created at rest). In
the medium, however, all nucleon states with momentum
$0\le |{\bf k}| \le k_{F}$ should be Pauli blocked and, hence, should
not contribute to the width. The threshold for pair production in the
medium should, therefore, move (for ${\bf q}=0$) from $q^{0}=2M^{*}$
to $q^{0}=2\sqrt{k_{F}^{2}+M^{*2}}$. This is precisely the region
\begin{equation}
  [q^{0}]_{\rm min} = 2M^{*} \le q^{0} \le 2\sqrt{k_{F}^{2}+M^{*2}}
    = [q^{0}]_{\rm max} \;,
\end{equation}
where the density-dependent contribution to the polarization
($\Pi^{22}_{D}$) develops an imaginary part which would exactly
cancel the contribution from vacuum polarization to the decay width of
the $\omega$ meson. Therefore, vacuum polarization must be included in
the study of the damping of the meson (collective) modes. The other
feature that one should stress are the many meson branches that are
developed for $0.3 \sim k_{F}/k_{F}^{0} \sim 0.9$. In
Ref.~\cite{limhor88} it has been suggested that this structure is also
related to the Pauli blocking of $N\bar{N}$ excitations. Indeed, by
removing the negative-energy (antiparticle) components from the
Feynman propagator it was shown that the multi-branch structure
disappears. Hence, vacuum polarization might also play an important
role in the modifications of the real part of the propagator and
should be included in the study of the effective $\omega$-meson mass.

We now present results for the effective $\omega$-meson mass in the
relativistic Hartree approximation (RHA). In the RHA one must include
vacuum contributions to the nucleon self-energy in the calculation of
the mean-field ground state as well as the vacuum dressing of the
meson propagator. Details of the renormalization procedure can be
found in Ref.~\cite{furhor88} (note that in the present work the
renormalization point is taken at $q_{\mu}^{2}=m_{v}^{2}$ and not
$q_{\mu}^{2}=0$). In Fig.~\ref{figtwo} we present results for the
effective $\omega$-meson mass obtained by finding the zeroes of the
inverse transverse propagator in analogy to the MFT case [see
Eq.~(\ref{dtrans})]. In the present RHA case, however, the
$\omega$-meson self-energy includes density-dependent corrections and
vacuum polarization. We observe, in particular, that the multi-branch
structure present in the MFT calculation has now completely
disappeared. Hence the effective mass in the medium (driven by virtual
excitations) as well as the modified width (driven by the decay into
$N\bar{N}$ pairs) are dramatically affected by vacuum polarization.
Note, however, that the combined effect of the density-dependent
dressing plus vacuum polarization generates, in contrast to the MFT
case, a reduction in the value of the effective $\omega$-meson mass in
the medium. One can shed some light into this result by studying the
low-density limit of vacuum polarization. In particular, if only
vacuum polarization is included in the dressing of the propagator one
obtains an effective meson mass given by
\begin{eqnarray}
  m_{v}^{*2} &=& m_{v}^{2} \left[ 1 + {g_{v}^{2} \over 3\pi^{2}}
    {(M^{*}\!-\!M) \over M} \right] + {\cal O}(m_{v}^{2}/4M^{2})
    \nonumber\\
  &=& m_{v}^{2} - \Omega^{2} {g_{v}^{2} \over 3\pi^{2}}
    {C_{s}^{2} \over C_{v}^{2}} + {\cal O}(m_{v}^{2}/4M^{2}) \;.
\end{eqnarray}
This result indicates that the shift in the mass of the $\omega$ meson
is proportional to the shift of the nucleon mass in the medium, and
thus negative in the Walecka model. If we now add both effects,
namely, the density-dependent dressing plus vacuum polarization, into
the calculation of the fully dressed propagator, we obtain (the
low-density limit of) the in-medium $\omega$-meson mass in the RHA:
\begin{equation}
  m_{v}^{*2} \simeq m_{v}^{2} + \Omega^{2} \left[ 1 -
    {g_{v}^{2} \over 3\pi^{2}} {C_{s}^{2} \over C_{v}^{2}} \right] \;.
  \label{mvrha}
\end{equation}
The above equation embodies the central result from the present work.
It indicates that the shift in the value of the $\omega$-meson mass
arises form a delicate competition between two effects. On the one
hand, the density-dependent dressing of the $\omega$-meson propagator
leads to an increase in the mass proportional to the classical plasma
frequency [see Eq.~(\ref{mvmft})]. On the other hand, the vacuum
dressing of the propagator is proportional to $M^{*}-M$ and gives
rise, in the present model, to the opposite effect. Because in the
Walecka model the scalar field is responsible for a downward shift in
the value of the nucleon mass in the medium, the $\omega$ meson
``drags'' along lighter $N\bar{N}$ pairs (relative to free space) that
are, ultimately, responsible for reducing the $\omega$-meson mass. The
final outcome of this competition depends on the particular values of
the coupling constants adopted in the model [see Eq.~(\ref{mvrha})].
In the present model
$(g_{v}^{2}/3\pi^{2})(C_{s}^{2}/C_{v}^{2})\sim 5.36 > 1$ and the
vacuum polarization dominates over the corresponding
density-dependent dressing leading to a reduction of the
$\omega$-meson mass in the medium. In the Walecka model, the values of
$C_{s}^{2}$ and $C_{v}^{2}$ were selected in order to reproduce, at
the mean-field level, the binding energy and density of nuclear matter
at saturation. By further associating the value of the vector-meson
mass to the physical value of $\omega$-meson mass, one obtains the
$NN\omega$ coupling constant given in Table~\ref{tbl.para}. One should
mention that this large value for the coupling constant is consistent
with other estimates based on fits to empirical two-nucleon
data\cite{machl87}. Thus we believe that the parameters adopted in
the present calculation are realistic, and that the shift in the value
of the $\omega$ meson should be largely insensitive to a fine tuning
of parameters. Other models, where the parameters have been
constrained by different physical observables, might (and actually
have) lead to different predictions in the magnitude, and even in the
direction, of the shift of the $\omega$-meson
mass\cite{hosaka90,hatlee92,koike93}. Therefore, the value of the
$\omega$-meson mass in the nuclear medium is model dependent and
should be determined experimentally\cite{cebaf89}.

We now close with a brief discussion of the longitudinal meson
propagator. In studying the longitudinal effective mass of the
$\omega$ meson in the medium one must find the zeroes of the
longitudinal propagator. Because of scalar-vector mixing, however,
Dyson's equation for the longitudinal propagator becomes a $3\times3$
matrix equation (actually a $2\times2$ matrix equation because of
current conservation). However our calculations, as well as those of
Ref.~\cite{limhor88}, suggest that the scalar-vector mixing is quite
small; the shift in the value of the $\omega$-meson mass is at most
1.5~MeV for all densities below nuclear-matter saturation density. We
have also shown that this result is fairly insensitive to the
particular choices of width and renormalization point for the scalar
propagator. In Fig.~\ref{figthree}, the longitudinal effective mass is
plotted as a function of the Fermi momentum for $|{\bf q}|=1$~MeV
(dashed line) and 1~GeV (dot-dashed line) along with the transverse
mass (solid line). For small values of $|{\bf q}|$, the transverse and
longitudinal effective masses are practically identical (in fact, they
are not resolved in the figure). This can be traced to the low-density
behavior of the longitudinal mass (ignoring scalar-vector mixing)
\begin{equation}
  m_{v}^{*2} \simeq m_{v}^{2} + \Omega^{2}
    \left[ {1\over1+{\bf q}^{2}/m_{v}^{2}} - {g_{v}^{2}\over 3\pi^{2}}
    {C_{s}^{2} \over C_{v}^{2}} \right] \;,
\end{equation}
which should be compared to Eq.~(\ref{mvrha}). For large values of the
momentum, however, the longitudinal effective mass is more sensitive
to the value of $|{\bf q}|$ and displays an even stronger reduction
(see Table~\ref{tbltwo}).

\section{Conclusion and Future Work}
\label{secconcl}

We have calculated the effective mass of the $\omega$ meson in nuclear
matter in the Walecka model. The ground state of nuclear matter was
obtained by solving the field equations in a mean-field approximation.
The effective mass of the $\omega$ meson was subsequently obtained by
solving Dyson's equation for the propagator in a relativistic
random-phase approximation. We have calculated the dressing of the
meson propagator in two approximations. In the MFT one neglects vacuum
polarization and computes the self-energy corrections to the
propagator by including the coupling of the meson to particle-hole
pairs and the Pauli blocking of $N\bar{N}$ excitations. In the RHA, on
the other hand, one computes the full (one-loop) dressing of the
propagator by also including vacuum polarization.

Both approximations have been used extensively in the past (with
considerable success) in the study of the linear response of the
nuclear system to a variety of probes. In all these cases the momentum
transfer to the nucleus, and hence the momentum carried by the meson,
was constrained to the spacelike region. In contrast to these
findings, we have shown that in the timelike region probed in the
present work, a MFT description is inappropriate. Because in the MFT
one includes the Pauli blocking of $N\bar{N}$ excitations, but not
vacuum polarization, one obtains an unphysical (i.e., negative)
contribution to the decay width of the $\omega$ meson. Furthermore,
one generates a dispersion curve for the collective meson modes having
a complicated multi-branch structure that arises from the
Pauli-blocked $N\bar{N}$ excitations.

These two obvious deficiencies of the MFT approach were corrected by
calculating the meson propagator in the RHA. In contrast to the MFT
results, the $\omega$-meson mass displayed a very smooth behavior as a
function of nuclear density. In addition, the damping of the meson
modes was now caused by the decay of the $\omega$ meson into
$N\bar{N}$ pairs. The Pauli blocking of $N\bar{N}$ excitations
(already present in the MFT description) simply reduced the decay
width of the $\omega$ meson by suppressing those transitions to
nucleon states below the Fermi surface.

In the RHA the effective $\omega$-meson mass was reduced relative to
its free-space value. This reduction arose from two competing effects.
On the one hand, the density-dependent dressing of the meson
propagator (with no vacuum polarization) caused an increase in the
$\omega$-meson mass proportional to the classical plasma frequency.
Vacuum polarization, on the other hand, lead to a reduction in the
mass. This reduction was, ultimately, traced to the corresponding
reduction of the nucleon mass in the medium. For the particular values
adopted in the model, vacuum polarization effects dominated over the
density-dependent dressing and lead to a reduction of about 170~MeV in
the value of the $\omega$-meson mass at nuclear-matter saturation
density (see Table~\ref{tbltwo}). We argue, that because the
parameters of the model were constrained by bulk properties of nuclear
matter at saturation, our finding should be largely insensitive to a
fine tuning of parameters. However, since other theoretical models,
constrained by a different set of observables, can apparently lead to
different predictions, it is important to make an experimental
determination of the effective $\omega$-meson mass.

In the future we plan to examine the effects of a medium-modified
$\omega$ mass to the photoproduction of $e^{+}e^{-}$ pairs in a
kinematical region around the $\omega$-meson mass pole. Many dynamical
components must be integrated in such a calculation. For example, in
addition to the calculation of an effective $\omega$-meson propagator
in the nuclear medium, one needs a model for the
$\gamma N \rightarrow \omega N$ transition amplitude that can be
extrapolated off-shell. Ideally, one would like to have a single
underlying model for the calculation of, both, the dressing of the
propagator and the photoproduction amplitude. One should also further
study the effect of the three-momentum transfer on the photoproduction
cross section. In free space the $\omega$-meson propagator is a
function of only one kinematical variable, namely, the four-momentum
squared ($q_{\mu}^{2}$). In the nuclear medium, the self-energy (and
thus the propagator) depends, in addition, on the magnitude of the
three-momentum transfer (${\bf q}$). For small values of the
three-momentum transfer the effective longitudinal and transverse
masses of the $\omega$ meson are practically identical. At larger
values of the three-momentum transfer ($|{\bf q}|\sim 1$~GeV),
however, the two modes get well separated with the longitudinal mass
reduced, near nuclear-matter saturation density, by an additional
50~MeV relative to the transverse value. This suggests that at large
enough values of the three-momentum transfer the spectrum for the
production of $e^{+}e^{-}$ pairs might show two well separated
($\omega$-meson) peaks. This statement obviously requires that the
signal be extracted from the wide ``backgrounds'' associated with
Bethe-Heitler (non-resonant) pairs and the formation and decay of the
(isovector) $\rho$ meson. This may be done by a careful study of the
interference of these amplitudes, for example\cite{cebaf89}. The study
of a medium-modified $\rho$-meson propagator is, unfortunately, much
more complicated in the context of a renormalizable quantum field
theory and will not be addressed here; see Ref.~\cite{serwal86}. We
believe that detailed measurements of the photoproduction of lepton
pairs from nuclei are essential and should provide invaluable
insights into the formation, propagation and decay of vector mesons
inside the nuclear medium.

\acknowledgments
This research was supported by the Florida State University
Supercomputer Computations Research Institute and U.S. Department of
Energy contracts DE-FC05-85ER250000, DE-FG05-92ER40750, and
DE-FG05-86ER40273. The research of AGW was supported in part by the
Australian Research Council.

\mediumtext
\begin{table}
 \caption{Mean-field parameters in the Walecka model. The nucleon mass
          and the $\omega$-meson mass were fixed at their physical
          values ($M=939$~MeV, $m_{v}=783$~MeV). The effective nucleon
          mass $M^{*}$ is the appropriate value at nuclear-matter
          saturation density ($k_{F}=1.3$~fm$^{-1}$).}
   \begin{tabular}{cdccccc}
      Model & $g_s^2$     &   $g_v^2$    & $m_s$(MeV) & $C_{s}^{2}$
            & $C_{v}^{2}$ &   $M^{*}/M$  \\
     \tableline
      MFT & 109.626 & 190.431 & 520 & 357.469 & 273.871 & 0.541  \\
      RHA &  54.289 & 102.770 & 458~& 228.198 & 147.800 & 0.730  \\
   \end{tabular}
 \label{tbl.para}
\end{table}

\narrowtext
\begin{table}
 \caption{Effective $\omega$-meson mass as a function of momentum
          transfer at nuclear-matter saturation density
          ($k_{F}=1.3$~fm$^{-1}$) in the RHA. Meson masses and the
          momentum transfer are measured in MeV.}
   \begin{tabular}{cdccc}
      Mode & $|{\bf q}|$   &   $m_{v}^{*}/m_{v}$   &   $m_{v}^{*}$
           & $m_{v}-m_{v}^{*}$ \\
     \tableline
      Transverse   & 1    & 0.786 & 615.7 & 167 \\
      Transverse   & 500  & 0.780 & 611.0 & 172 \\
      Transverse   & 1000 & 0.776 & 607.5 & 176 \\
      Longitudinal & 1    & 0.786 & 615.7 & 167 \\
      Longitudinal & 1000 & 0.718 & 562.1 & 221 \\
   \end{tabular}
 \label{tbltwo}
\end{table}

\begin{figure}
\caption{Effective transverse $\omega$-meson mass as a function of the
        Fermi momentum for $|{\bf q}|=1$~MeV and $|{\bf q}|=1$~GeV in
        MFT. The dotted lines enclose the ($q^{0}$) region where the
        imaginary part of the transverse polarization is nonzero and
        damping of the modes is possible.}
 \label{figone}
\end{figure}

\begin{figure}
\caption{Effective transverse $\omega$-meson mass as a function of the
        Fermi momentum for $|{\bf q}|=1$, 500, and 1000~MeV in RHA.
        The damping of the modes occurs for values of $q^{0}$ outside
        the range shown in the figure.}
 \label{figtwo}
\end{figure}

\begin{figure}
\caption{Effective transverse $\omega$-meson mass as a function of the
        Fermi momentum for $|{\bf q}|=1$~MeV (solid line) in RHA. Also
        shown is the longitudinal mass at $|{\bf q}|=1$~MeV (dashes)
        and $|{\bf q}|=1$~GeV (dot-dashed). The damping of the modes
        occurs for values of $q^{0}$ outside the range shown in the
        figure. Note that the transverse and longitudinal modes at
        1~MeV are indistinguishable in this figure.}
 \label{figthree}
\end{figure}

\end{document}